\documentclass[3p,times,twocolumn]{elsarticle}
 \biboptions{comma,sort&compress}
 
\usepackage{graphicx}
\usepackage{amsmath}
\usepackage{here}
\usepackage{subcaption}
\usepackage{hyperref}
\usepackage{nccmath}
\usepackage{booktabs}
\usepackage{textgreek}
\usepackage{lineno}
\usepackage{ecrc}

\volume{00}

\firstpage{1}

\journalname{Nuclear and Particle Physics Proceedings}
\runauth{Luca Quaglia}

\jid{nppp}
\jnltitlelogo{Nuclear and Particle Physics Proceedings}

\usepackage{amssymb}

\usepackage[figuresright]{rotating}

\begin{document}

\begin{frontmatter}

\title{
%
Recent results and upgrade of the ALICE muon spectrometer\,$^*$} 
 
 \cortext[cor0]{Review talk presented at QCD24, 27th High-Energy Physics International Conference in Quantum Chromodynamics 8-12/07/2024, Montpellier - FR.}

 \author[label1]{Luca Quaglia}
\address[label1]{Istituto Nazionale di Fisica Nucleare (INFN), Sezione di Torino, via Pietro Giuria 1, 10125, Torino, Italy}
\ead{luca.quaglia@to.infn.it}
\ead{luca.quaglia@cern.ch}

\author[]{on behalf of the ALICE collaboration}

\pagestyle{myheadings}
\markright{ }
\begin{abstract}
\noindent
The ALICE experiment at the CERN Large Hadron Collider (LHC) is a multi-purpose particle detector, mainly focused on the study of quark-gluon plasma (QGP) in heavy-ion collisions. In the forward rapidity region, 2.5 $<$ y $<$ 4, ALICE is equipped with a muon spectrometer (MS), which allows to study quarkonia and open heavy-flavor particles, both key probes to investigate QGP properties.

Although in LHC Run 1 and 2 many important results were achieved, the front absorber of the MS represented a limit to the physics program, due to the multiple scattering and energy loss in the material. To assess this limitation, a new forward vertex tracker (Muon Forward Tracker, MFT) was installed between the inner tracking system (ITS) and the front absorber. This has enhanced the MS physics performance, enabling the separation of prompt/non-prompt charmonium production at forward rapidity. It will also allow one to reduce the combinatorial background from semi-leptonic decays of kaons and pions. Finally, it will greatly improve the invariant-mass resolution of the low-mass dimuon pairs.

Moreover, during the ongoing LHC Run 3, the rate of Pb\textendash{}Pb collisions has been increased from 10~kHz (in Run 2) up to 50~kHz, allowing to collect a data sample about 5 times larger than the one recorded in Run 2.

This contribution will provide a brief overview of the MS upgrades and it will focus on the expected physics performance during the LHC Run 3. Some of the preliminary results already obtained will also be shown.
 
\end{abstract}
\end{frontmatter}
\section{Introduction}
\label{sec:intro}
A Large Ion Collider Experiment (ALICE) \cite{aliceTDR}, is one of the four large experiments located along the Large Hadron Collider (LHC) \cite{LHC}. ALICE is a multi-purposed particle detector, taking data in all LHC colliding systems (i.e. proton-proton (pp), lead-lead (Pb\textendash{}Pb), and proton-lead (p-Pb)), which mainly focuses on the study of the quark-gluon-plasma (QGP) in heavy-ion collisions.

QGP is a state of matter, where quarks and gluons are deconfined and can move freely over distances larger than the hadron size. This state of matter is created in ultra-relativistic heavy-ion collisions when critical temperature and energy densities are reached \cite{QGP1,QGP2}. QGP itself is short-living ($\approx$10~fm/\textit{c}) so it cannot be observed directly. For this reason, the study of particles (namely hard probes) which experience the evolution of the strongly interacting medium produced in heavy-ion collisions plays a crucial role. A few examples of QGP probes, studied in the ALICE MS include: quarkonium suppression, J/$\psi$ flow, and heavy-flavor production. These probes are studied through the muonic decay channel, as better described in the following. 

\begin{figure}
\includegraphics[width=\linewidth]{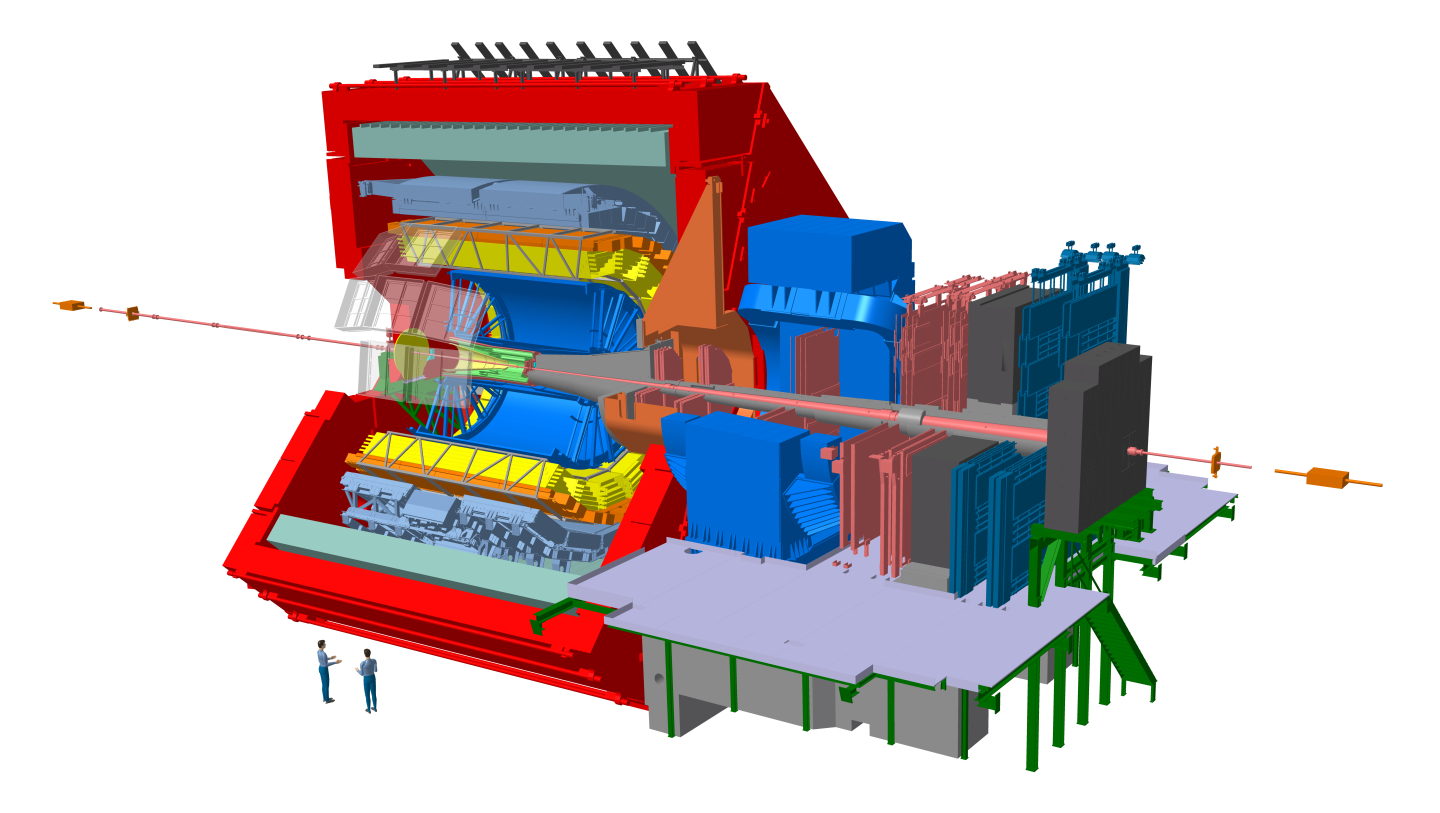}
\caption{Sketch of the ALICE detectors as of LHC Run 3 (from 2021). The central barrel detectors are located inside of the red solenoid magnet and the muon spectrometer is in the forward rapidity region.}
\label{fig:alice}
\end{figure}

Figure \ref{fig:alice} shows a scheme of the ALICE detectors, as of the LHC Run 3 (2021 onward). The apparatus is divided into two spatially distinct detection regions: the \textit{central barrel}, which covers the pseudorapidity region $|\eta| <$ 0.9 and  provides particle tracking and identification, and the \textit{muon spectrometer} (MS) which covers the pseudorapidity range -4 $<$ $\eta$ $<$ -2.5, that provides muon tracking and dimuon reconstruction down to $\it{p}_{\rm{T}}$ = 0.

This contribution is divided as follows: Section \ref{sec:run2} contains a description of the MS, together with a few highlights of its physics performance up to LHC Run 2, along with its limitations and motivations for upgrade; Section \ref{sec:run3} describes the upgrade of the MS; Section \ref{sec:run3Potential} contains a report of the expected physics performance, and some preliminary results obtained in pp and Pb\textendash{}Pb collisions during LHC Run 3 are described in Section \ref{sec:results}. Lastly, Section \ref{sec:conclusion} contains a summary of the results and the closing remarks.

\section{The ALICE MS up to LHC Run 2}
\label{sec:run2}

This section describes the layout of the ALICE MS up to LHC Run 2 (in \ref{sub:layoutRun2}) and a few highlights of its physics performances (in \ref{sub:perfRun2}).

\subsection{Muon Spectrometer layout in Run 2}
\label{sub:layoutRun2}

\begin{figure}[h] 
\centering
\includegraphics[width=0.85\linewidth]{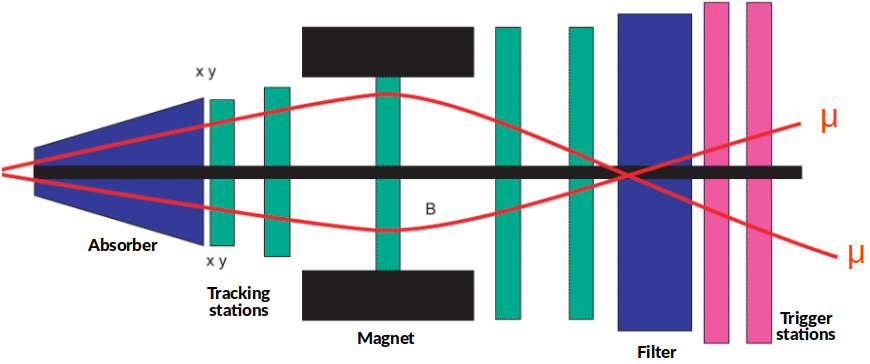}
\caption{Sketch of the ALICE MS up to the end of LHC Run 2 (a detailed description of its components is reported in the text).}
\label{fig:msRun2}
\end{figure}

Figure \ref{fig:msRun2} shows a sketch of the MS up to the end of LHC Run 2 (in 2018). 

Its components are listed here and for a more detailed description the reader can refer to \cite{muonTDR}:

\begin{itemize}
    \item \textbf{Front absorber} and \textbf{hadron filter}: composite concrete cone (before the tracking chambers) and iron wall (between the tracking and trigger chambers) to filter hadron contamination .

    \item \textbf{Dipole magnet}: providing a 3.5 T$\cdot$m integrated magnetic field, to bend muon tracks and measure their charge and momentum.

    \item \textbf{Tracking stations}: 5 planes of cathode pad/strip chambers to provide muon tracking (internally referred to as \textit{Muon CHambers}/\textit{MCH}).

    \item \textbf{Trigger stations}: 4 planes of Resistive Plate chambers, to provide a trigger signal to the MS (internally referred to as \textit{Muon TRigger}/\textit{MTR}).
\end{itemize}

As it was already anticipated in Section \ref{sec:intro}, the MS exploits the muonic decays of quarkonia and open heavy-flavors to study the QGP properties. In particular, to achieve our physics goals, the system must be capable of studying quarkonia down to $\it{p}_{\rm{T}}$ = 0 (since their production is sensitive to QGP effects that are large at low $\it{p}_{\rm{T}}$) as well as to disentangle the 3 $\Upsilon$ states (1S, 2S, and 3S), and, for this reason, it requires an invariant mass resolution of $\approx$150~MeV/\textit{c}$^{2}$ and $\approx$70~MeV/\textit{c}$^{2}$ for the $\Upsilon$ and J/$\psi$, respectively \cite{muonTDR}.

\subsection{Physics performance in Run 2}
\label{sub:perfRun2}

Heavy quarks are produced in the first stages of the collisions, hence they experience the full evolution of the system, losing energy when interacting with the medium, mainly with radiative and collisional energy loss.

In order to study the in-medium effects, due to the presence of QGP, one can introduce the nuclear modification factor ($\it{R}_{\rm{AA}}$), defined in Eq. \ref{eq:raa}:

\begin{equation}
    \centering
    \label{eq:raa}
    \it{R}_{\text{AA}} = \frac{1}{<T_{\text{AA}}>} \frac{d^{2}N/dp_{\text{T}}dy}{d^{2}\sigma_{\text{pp}}/dp_{\text{T}}dy}
\end{equation}

where $<T_{\text{AA}}>$ is the average nuclear overlap function, obtained with a Glauber Monte Carlo simulation \cite{glauber}, $d^{2}N/dp_{\text{T}}dy$ is the $\it{p}_{\rm{T}}$- and $y$-differential particle invariant yield in Pb\textendash{}Pb collisions and $d^{2}\sigma_{\text{pp}} / dp_{\text{T}}dy$ the $\it{p}_{\rm{T}}$- and $y$-differential production cross section in pp collisions at the same center-of-mass energy as the Pb\textendash{}Pb collisions. The $\it{R}_{\text{AA}}$ is a measure of the binary-scaling breaking in Pb\textendash{}Pb collisions, with respect to pp ones.

\subsubsection{Quarkonium production}
\label{subsub:supprRun2}

\begin{figure}[h] 
\centering
\includegraphics[width=0.85\linewidth]{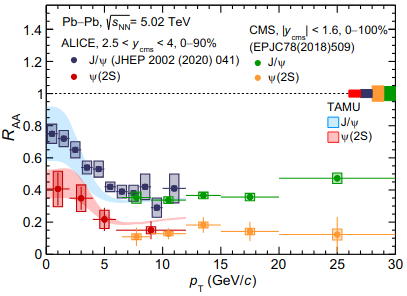}
\caption{Nuclear modification factor ($\it{R}_{\text{AA}}$) for the J/$\psi$ and $\psi$(2S) as a function of their $\it{p}_{\rm{T}}$ in Pb\textendash{}Pb collisions at $\sqrt{s_{\text{NN}}}$ = 5.02~TeV at the LHC \cite{quarkoniaSuppr}.}
\label{fig:jPsiSupprRun2}
\end{figure}

A large density of free colour charges is present in the QGP, leading to a screening of the q$\bar{\text{q}}$ binding potential \cite{colorScreen}, with a consequent dissociation of the bound states. Charmonia (c$\bar{\text{c}}$) exist in different states, for instance J/$\psi$ (ground state) and $\psi$(2S) (excited state). Figure \ref{fig:jPsiSupprRun2} shows the $\it{R}_{\text{AA}}$ for both resonances, as a function of their $\it{p}_{\rm{T}}$ in Pb\textendash{}Pb collisions at $\sqrt{s_{\text{NN}}}$ = 5.02~TeV.

Due to their different binding energies ($\approx$640 and $\approx$50~MeV for the J/$\psi$ and $\psi$(2S) respectively \cite{bindEn}), the two resonances are suppressed at different temperatures, which also leads to the clear $\it{R}_{\text{AA}}$ hierarchy observed in Fig. \ref{fig:jPsiSupprRun2}. The $\it{R}_{\text{AA}}$ increase, observed at low $\it{p}_{\rm{T}}$ can be explained by the recombination of uncorrelated c$\bar{\text{c}}$ pairs in the QGP \cite{recombination}. Moreover, Fig. \ref{fig:jPsiSupprRun2} also shows data from CMS \cite{cmsRaa}, at higher $\it{p}_{\rm{T}}$, which are similar to those provided by ALICE at lower $\it{p}_{\rm{T}}$. It is also worth mentioning that ALICE has measured the J/$\psi$ $\it{R}_{\text{AA}}$ in Pb\textendash{}Pb collisions for higher $\it{p}_{\rm{T}}$ values and with better precision as reported in \cite{higPtJPsiALICE}.

\subsubsection{J/\texorpdfstring{$\psi$}{psi} flow in Pb\textendash{}Pb collisions}
\label{subsub:flowRun2}

\begin{figure}[h] 
\includegraphics[width=\linewidth]{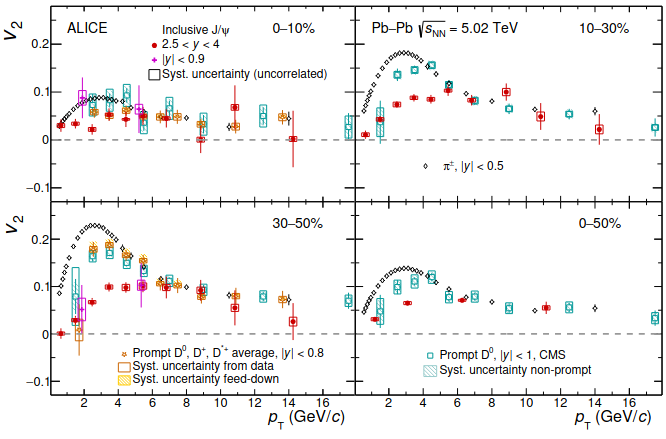}
\caption{$v_{2}$ coefficient for inclusive J/$\psi$ at forward and midrapidity, prompt D mesons at midrapidity and $\pi^{\pm}$ from ALICE, and prompt D$^{0}$ at midrapidity from CMS, as a function of $\it{p}_{\rm{T}}$  in Pb\textendash{}Pb collisions at $\sqrt{s_{\text{NN}}}$ = 5.02~TeV at the LHC for various centrality classes \cite{flow}.}
\label{fig:flow}
\end{figure}

The azimuthal dependence of particle production, in heavy-ion collisions, can be quantified by means of a Fourier expansion in terms of the difference between the azimuthal angle of the particle ($\Phi$) and the angle of the initial symmetry plane ($\Psi_{n}$) \cite{flow}, as reported in Eq. \ref{eq:flow}:

\begin{equation}
    \centering
    \label{eq:flow}
    \frac{dN}{d\Phi} = 1 + 2 \sum_{n=1}^{\infty} v_{n} \cos [n(\Phi - \Psi_{n})]
\end{equation}

where $v_{n}$ is the n-th order harmonic coefficient of the Fourier expansion. The largest contribution is given by the second order harmonic ($v_{2}$), also known as \textit{elliptic flow} \cite{flow} and it is caused by the ellipsoidal shape of the overlap region in non-central heavy-ion collisions. This initial spatial asymmetry is translated to a momentum anisotropy of the final-state particles. Figure \ref{fig:flow} shows the values of $v_{2}$, calculated with the scalar-product method \cite{scalarProduct}, as a function of $\it{p}_{\rm{T}}$ in Pb\textendash{}Pb collisions at $\sqrt{s_{\text{NN}}}$ = 5.02~TeV for various centrality classes. 

The four panels of Fig. \ref{fig:flow} show the values of $v_{2}$ for: inclusive (prompt and non-prompt) J/$\psi$ at forward and midrapidity (in red and magenta, respectively), prompt D mesons at midrapidity (in yellow), $\pi^{\pm}$ (in black)  as well as for prompt D$^{0}$ mesons at midrapidity measured by CMS \cite{cmsFlow} in green.

The $v_{2}$ values increase with increasing $\it{p}_{\rm{T}}$ up to a maximum for intermediate $\it{p}_{\rm{T}}$ values ($\approx$4~GeV/\textit{c}) and for more peripheral collisions. Moreover, the $v_{2}$ values are also found to be statistically compatible between forward and midrapidity. Lastly, there is a particle mass ordering of the $v_{2}$ values for $\it{p}_{\rm{T}}<$ 6~GeV/\textit{c} while for $\it{p}_{\rm{T}}>$ 8~GeV/\textit{c} the $v_{2}$ measurements converge to similar values.

\subsubsection{Heavy-flavor production}
\label{subsub:heavyProdRun2}

\begin{figure}[h!] 
\centering
\includegraphics[width=\linewidth]{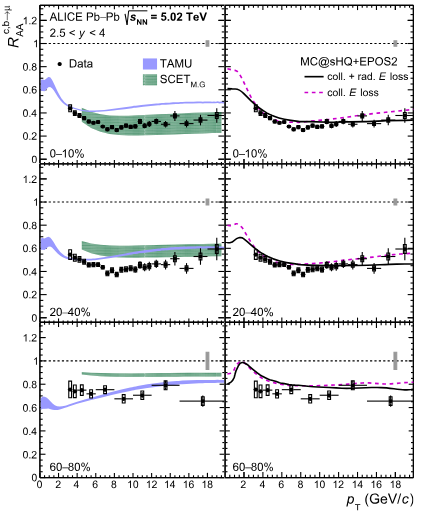}
\caption{Inclusive $\it{R}_{\text{AA}}$ for muons from b and c quarks as a function of $\it{p}_{\rm{T}}$ in Pb\textendash{}Pb collisions at $\sqrt{s_{\text{NN}}}$ = 5.02~TeV at the LHC \cite{heavyFlav}. Left panels: comparison of data with TAMU and SCET models. Right panels: comparison with the MC@sHQ+EPOS2 model.}
\label{fig:heavyProd}
\end{figure}

Inclusive heavy-flavor (particles containing either a b or c quark, together with a light one) production is studied in the MS. Figure \ref{fig:heavyProd} shows the inclusive $\it{R}_{\text{AA}}$ for b and c quarks, as a function of $\it{p}_{\rm{T}}$ in various centrality classes.

The suppression is larger in more central collisions, where it reaches a maximum in the range 6 $<$ $\it{p}_{\rm{T}}<$ 10~GeV/\textit{c} while, it is less pronounced in the more peripheral collisions. Another observation is that, in p-Pb collisions, where the formation of QGP is not expected, the R$_{\text{pPb}}$ is $\approx$1 \cite{heavyFlavpPb}, meaning that the suppression observed in Pb\textendash{}Pb collisions is due to final-state interactions between b and c quarks with the QGP.

Figure \ref{fig:heavyProd} also contains the comparison between the experimental data and some models. Specifically, the data is compared to the following models: TAMU \cite{TAMU}, which describes the interactions between quarks and QGP as purely elastic, SCET \cite{SCET}, which is a pQCD-based model that implements medium-induced gluon radiation by means of a modified splitting function and finite quark masses (left panels of Fig. \ref{fig:heavyProd}), and MC@sHQ+EPOS2 \cite{MC,hepos}, which contains a hydrodynamic description of the medium, coupled with different parton energy losses. Out of the three models, the latter is closer to the data than the others, over the whole $\it{p}_{\rm{T}}$ range and within uncertainties (right panels of Fig. \ref{fig:heavyProd}).

\section{The ALICE MS from Run 3 onward}
\label{sec:run3}

The results discussed in Section \ref{sub:perfRun2} have shown the capabilities of the MS although it also came with some limitations. Indeed, the large distance between the interaction vertex and the first tracking station, coupled with the track smearing due to the multiple scattering in the front absorber, makes it impossible to constrain the tracks in the primary vertex region. This leads to a limited rejection power for muons coming from the semi-leptonic decays of kaons and pions and to an inability to separate the prompt and non-prompt J/$\psi$ components. Moreover, multiple scattering in the absorbers leads to a degradation of the track's angular resolution, affecting especially the measurements at low dimuon invariant mass.

\begin{figure}[h] 
\includegraphics[width=\linewidth]{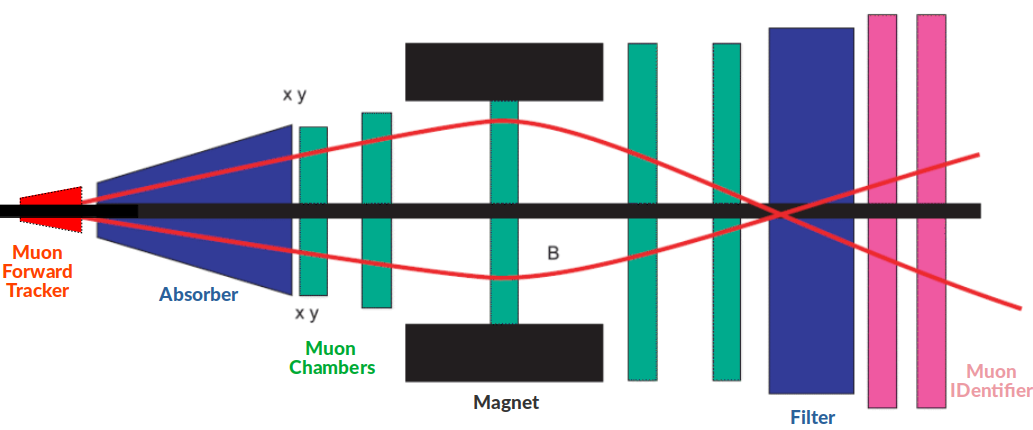}
\caption{Sketch of the ALICE MS from LHC Run 3 onward (a detailed description of its components is reported in the text).}
\label{fig:msRun3}
\end{figure}

These considerations, together with the higher luminosity and interaction rate (up to 6$\times$10$^{27}$~cm$^{-2}$s$^{-1}$ and 50 kHz, respectively, in Pb\textendash{}Pb collisions \cite{highRate}) foreseen for Run 3, lead to a massive upgrade of the whole ALICE detectors \cite{upgrades}. The MS was also part of this extensive upgrades and Fig. \ref{fig:msRun3} shows a sketch of the upgraded MS. The main upgrades regard the following points:

\begin{itemize}
    \item A new silicon pixel tracker, namely \textbf{Muon Forward Tracker} (MFT), made out of 5 pixel disks was installed in front of the hadron absorber. This detector enables muon tracking upstream of the absorber and, by matching the tracks in the MFT with those in the MCH, it enables tracking by the MS also in the primary-vertex region \cite{mftLOI,mftTDR}.

    \item The front-end electronics of the MCH have been upgraded to a new \textit{DUAL-SAMPA} \cite{dualSampa} chip and a new readout chain has been installed in order to work in continuous readout mode, to cope with the higher interaction rate.

    \item New front-end electronics (FEERIC \cite{feeric}) have been installed in the MTR RPCs and the readout electronics have also been adapted to work in continuous readout mode. Most notably, thanks to the continuous readout mode, no hardware trigger is needed in the MS anymore, hence the MTR has become a Muon IDentifier (\textbf{MID}).  
\end{itemize}

\section{ALICE MS physics potential from Run 3 onward}
\label{sec:run3Potential}

This section describes the new physics potential unlocked by the upgraded MS, using the results of simulations that have been implemented to characterize the upgraded MS.

\subsection{Prompt and non-prompt J/\texorpdfstring{$\psi$}{psi} at forward rapidity}
\label{sub:nonPromptJpsi}

B hadrons have non-negligible \cite{nonPrompt} branching ratios for their decays into a J/$\psi$ and other particles (indicated as \textit{X} in the following). The B hadrons decay length (\textit{c}$\tau$) is about 420-490~nm) and to identify these decays, a precise measurement of the particle production vertex is needed. 

\begin{figure}[h] 
\centering
\includegraphics[width=0.6\linewidth]{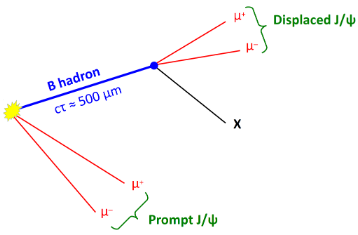}
\caption{Schematic representation of prompt and non-prompt (from B-hadron decay) J/$\psi$ production \cite{mftTDR}.}
\label{fig:displacedVertex}
\end{figure}

In the case of the ALICE MS, this is made possible thanks to the addition of the MFT. Figure \ref{fig:displacedVertex} shows a schematic representation of a B hadron decaying to a displaced J/$\psi$, together with a prompt J/$\psi$ produced at the primary interaction vertex.

As reported in \cite{mftTDR}, it is possible to study B-hadron down to $\it{p}_{\rm{T}}$ = 0, thanks to a peculiar kinematic correlation between the daughter J/$\psi$ and parent B-hadron: it can indeed be shown that a B-hadron with $\it{p}_{\rm{T}}$ = 0 will produce a J/$\psi$ with a $\it{p}_{\rm{T}}$ $\approx$ 1.5~GeV/\textit{c}, thanks to the presence of the above mentioned other decay products (X) and such a $\it{p}_{\rm{T}}$ value is well within the reach of the MS.

\begin{figure}[h]
    \centering
    \begin{subfigure}[t]{0.5\linewidth}
        \centering
        \includegraphics[height=1.19in]{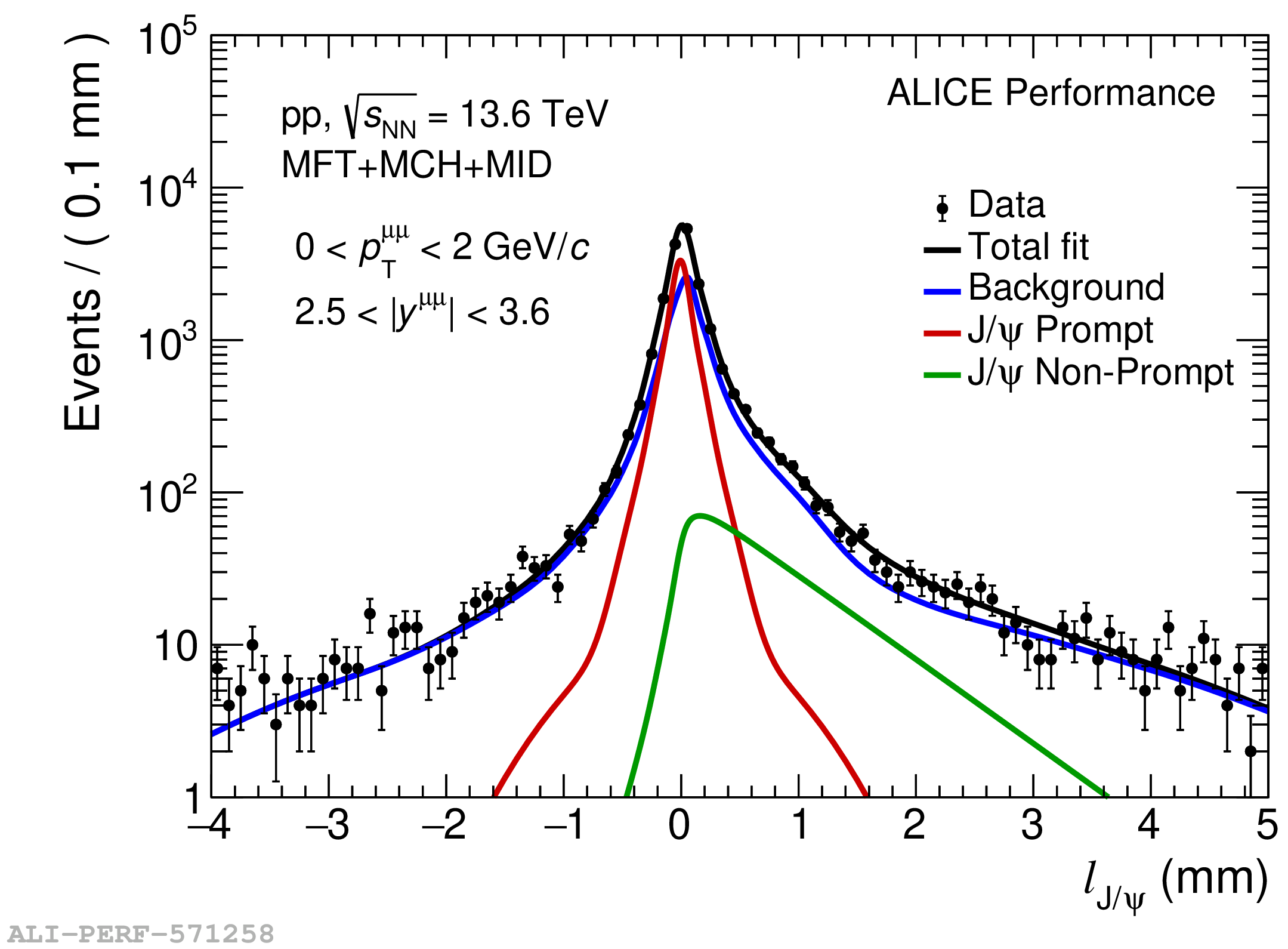}
        \caption{0 $<$ $\it{p}_{\rm{T}}$ $<$ 2 GeV/c}
        \label{fig:promptNonPromptPt1}
    \end{subfigure}%
    ~ 
    \begin{subfigure}[t]{0.5\linewidth}
        \centering
        \includegraphics[height=1.19in]{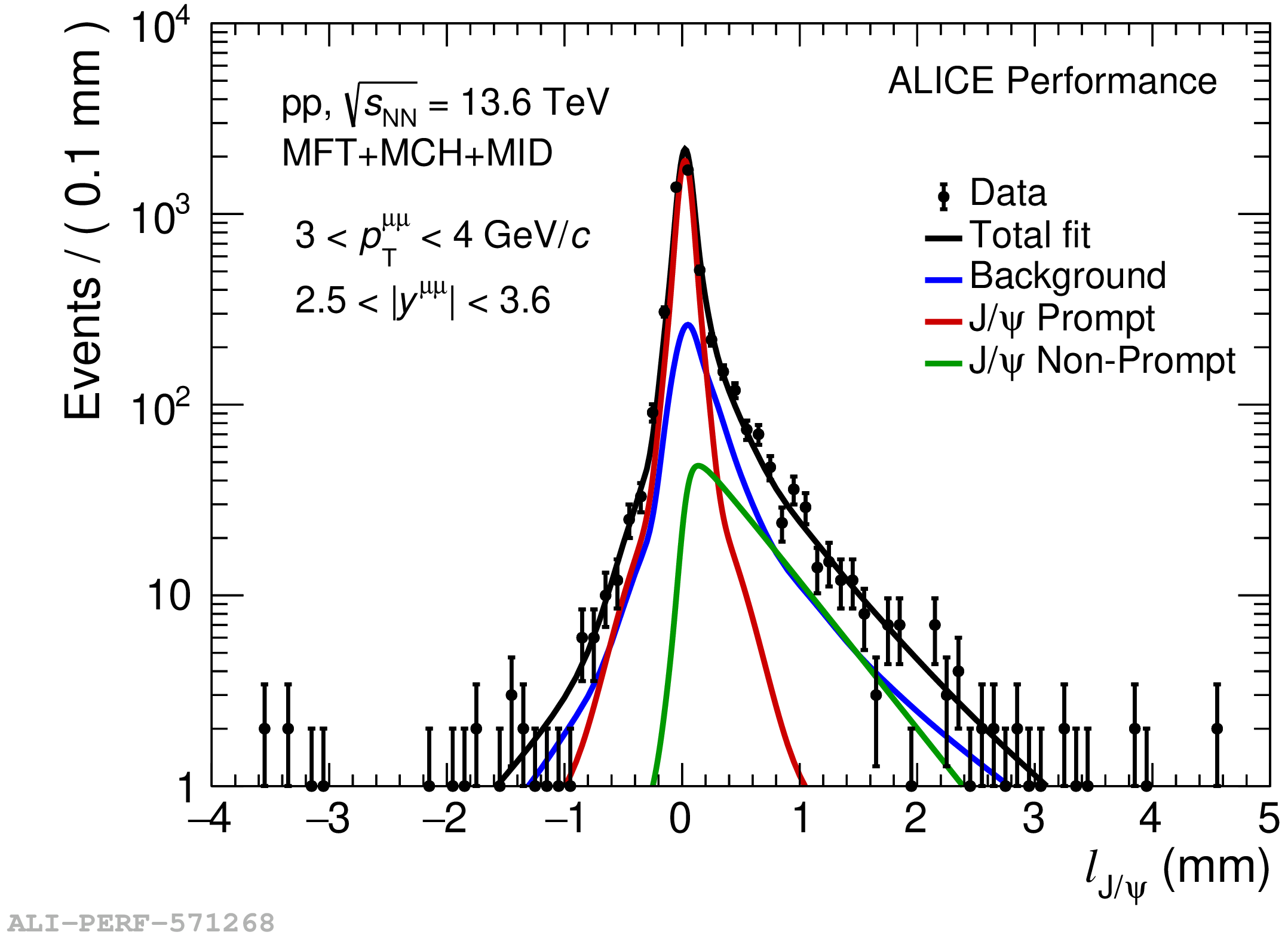}
        \caption{3 $<$ $\it{p}_{\rm{T}}$ $<$ 4 GeV/c}
        \label{fig:promptNonPromptPt2}
    \end{subfigure}
    \caption{Pseudo-proper decay length distribution with the fits for prompt/non-prompt J/$\psi$, and background components in pp collisions at $\sqrt{s}$ = 13.6~TeV.}
\end{figure}

With the addition of the MFT, it is possible to precisely measure the distance between the primary vertex and the decay vertex of all the detected particles, allowing one to statistically separate the prompt and non-prompt J/$\psi$ components according to the different distributions of their distance from the primary vertex (L). Assuming that $\gamma_{\text{J}/\psi} \approx \gamma_{B}$ (where $\gamma$ is the Lorentz factor for both particles), it is possible to introduce the \textit{pseudo-proper decay time} (\textit{t}) of the J/$\psi$ as reported in Eq. \ref{eq:pseudoProper}:

\begin{equation}
    \label{eq:pseudoProper}
    \hspace*{1.4cm}
    t = \frac{|\overrightarrow{r}_{\text{J}/\psi} - \overrightarrow{r}_{vtx} | \cdot M_{\text{J}/\psi}}{p} 
\end{equation}

where $\overrightarrow{r}_{J/\psi}$ and $\overrightarrow{r}_{vtx}$ represent the secondary and primary vertex position, $M_{\text{J}/\psi}$ the J/$\psi$ mass and $p$ its momentum. The pseudo-proper decay time can be decomposed in its transverse ($t_{xy}$) and longitudinal ($t_{z}$) components. In ALICE, $t_{xy}$ is used at midrapidity and $t_{z}$ at forward rapidity, as reported in Eq. \ref{eq:pseudoProperz}:

\begin{equation}
    \label{eq:pseudoProperz}
    \hspace*{1.4cm} 
    t_{z} = \frac{(z_{J/\psi} - z_{vtx}) \cdot M_{J/\psi}}{p_{z}}
\end{equation}

where the 3-dimensional primary, secondary-vertex positions and momentum have been replaced by their longitudinal components. The t$_{z}$ distribution is peaked around zero (for prompt J/$\psi$) while the displaced J/$\psi$ has a positive tail that reflects the B-hadron decay time. In addition, the t$_{z}$ distribution can have a contribution from uncorrelated dimuon background. By performing a fit to the invariant-mass spectrum and to the t$_{z}$ distribution (with a variable width Gaussian), it is possible to first fix the normalization parameters of the background and signal, in terms of inclusive J/$\psi$, from the invariant-mass spectrum and then to decompose the t$_{z}$ distribution in its components (prompt/non-prompt J/$\psi$ and background), leaving the prompt/non-prompt ratio as a free parameter. Figures \ref{fig:promptNonPromptPt1} and \ref{fig:promptNonPromptPt2} show the result of the fit just described for the pseudo-proper decay length (l$_{z}$ = \textit{c}$\cdot$t$_{z}$, where \textit{c} is the speed of light), for two different $\it{p}_{\rm{T}}$ intervals.

\begin{figure}[h] 
\centering
\includegraphics[width=0.6\linewidth]{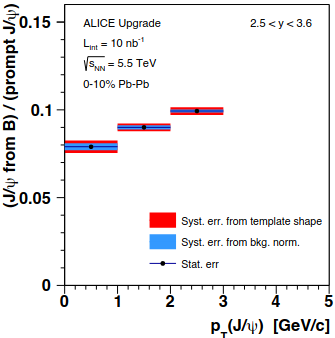}
\caption{Expected results in Run 3 (from simulation) for the prompt to non-prompt J/$\psi$ ratio measured at forward rapidity,  \cite{mftTDR}.}
\label{fig:ratio}
\end{figure}

The prompt/non-prompt J/$\psi$ components are clearly visible and they can be used to calculate the ratio between prompt and non-prompt J/$\psi$, which is shown as a function of $\it{p}_{\rm{T}}$ in Fig. \ref{fig:ratio}.

The main sources of uncertainty in this measurement are: 0.8$-$1.6\% statistical (from the t$_{z}$/l$_{J/\psi}$ fit), together with a 1\% systematic uncertainty from the background normalization and another from the shape of the l$_{z}$ distribution. In conclusion, an error below 5\% can be expected on the non-prompt to prompt J/$\psi$ fraction down to $\it{p}_{\rm{T}}$ = 0 \cite{mftTDR}. 

\subsubsection{Non-prompt J/\texorpdfstring{$\psi$}{psi} \texorpdfstring{$\it{R}_{\text{AA}}$}{RAA}}
\label{subsub:nonPromptJpsiRaa}

\begin{figure}[h] 
\centering
\includegraphics[width=0.6\linewidth]{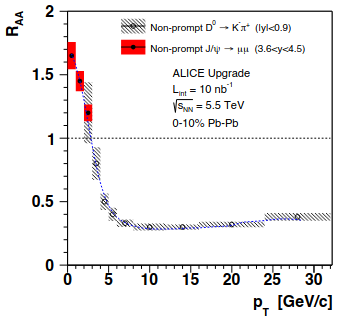}
\caption{Expected results in Run 3 (from simulation) for the non-prompt J/$\psi$ $\it{R}_{\text{AA}}$ measured at forward rapidity, \cite{mftTDR}.}
\label{fig:nonPromptJPsiRaa}
\end{figure}

Once the non-prompt to prompt J/$\psi$ fraction has been estimated, the non-prompt J/$\psi$ $\it{R}_{\text{AA}}$ can also be computed. To estimate the uncertainty on this quantity, the one coming from the pp reference (pp collisions at the same center-of-mass energy as the Pb\textendash{}Pb ones) measurement is needed. To this end, the signal obtained in Pb\textendash{}Pb is scaled to the expected pp-reference L$_{\text{int}}$, the statistical uncertainty is obtained by fitting the t$_{z}$ distribution in pp collisions (with a quasi-perfect matching between the MFT and MS due to the much lower background in this case) and the systematic one is assumed to be the same as in Pb\textendash{}Pb collisions. This allows one to measure the non-prompt J/$\psi$ $\it{R}_{\text{AA}}$ down to $\it{p}_{\rm{T}}$ = 0, with a total uncertainty $<$ 5\%, as it is reported in Fig. \ref{fig:nonPromptJPsiRaa}, together with the expected physics performance of the upgraded ALICE Inner Tracking System (ITS \cite{ITS}) in the measurement of beauty via non-prompt D$^{0}$ in the midrapidity region. It can be clearly seen how the measurements at forward rapidity will enhance the physics reach of ALICE.

\subsection{Open heavy-flavor via single-muon decay channel}
\label{sub:heavyFlavRun3}

The installation of the MFT will also contribute to the study of open heavy-flavors in the single-muon and dimuon channels. The latter needs to be performed down to very low $\it{p}_{\rm{T}}$, to extract the charm and beauty production cross section with the least model dependence as possible \cite{charmReg}.

It is also interesting to study beauty and charm production separately, in order to investigate their different energy loss in QGP and also the expected $\it{R}_{\text{AA}}$ hierarchy ($\it{R}_{\text{AA}}^{D} < \it{R}_{\text{AA}}^{B}$) \cite{hierarchy}. 

The results reported in the following have been obtained by means of a Monte Carlo simulation, including the MFT in the ALICE MS. The signal is defined as the \textit{b} or \textit{c} direct decays or the decay chain: $\mu \leftarrow$ D $\leftarrow$ B, while the background is defined as the muons coming from particle decays in the underlying simulated event. Since the decay of b and c is displaced, with respect to the primary vertex, the distribution of this offset is generated for the signal and background and they are fitted in each $\it{p}_{\rm{T}}$ interval with a variable width Gaussian. The global offset distribution is then built by adding up the open charm, open beauty and background distributions. This is then re-fitted with a template function, whose parameters are tuned on the Monte Carlo, with the normalization of each contribution left as free parameters \cite{mftTDR,mftLOI}. In general, the statistical uncertainty obtained is below 0.5\%. Four main sources of systematical uncertainties have been considered in \cite{mftLOI} and are briefly outline here:

\begin{itemize}
    \item Residual misalignment between ITS and MFT 
    \begin{itemize}
        \item ITS is used to measure the position of the primary vertex
        \item Primary vertex is moved by $\pm$10~$\mu$m and data re-fitted
    \end{itemize}

    \item Residual dependence of the average offset from the parent hadron $\it{p}_{\rm{T}}$

    \item MFT pointing resolution 
    \begin{itemize}
        \item Estimated with a Gaussian smearing of the reconstructed vertex
    \end{itemize}

    \item MS tracking efficiency
    \begin{itemize}
        \item Assumed to be $\approx$ 2.5\% at the track level
    \end{itemize}
\end{itemize}

\begin{figure}[h]
    \centering
    \begin{subfigure}[t]{0.5\linewidth}
        \centering
        \includegraphics[height=1.15in]{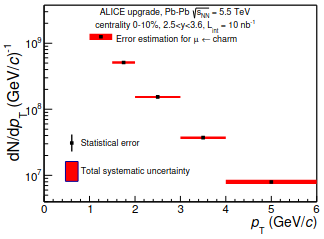}
        \caption{Open charm yield vs $\it{p}_{\rm{T}}$}
        \label{fig:openCharmRun3}
    \end{subfigure}%
    ~ 
    \begin{subfigure}[t]{0.5\linewidth}
        \centering
        \includegraphics[height=1.15in]{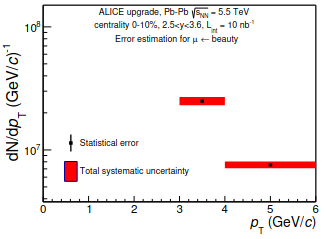}
        \caption{Open beauty yield vs $\it{p}_{\rm{T}}$}
        \label{fig:openBeautyRun3}
    \end{subfigure}
    \caption{Expected results in Run 3 (from simulation) for the $\it{p}_{\rm{T}}$-differential yield for open charm and beauty in the single-muon channel \cite{mftTDR}.}
\end{figure}

In general, the total uncertainty is $<$ 10\% for open charm with $\it{p}_{\rm{T}}$ = 1~GeV/c while for open beauty, measurement is robust only for $\it{p}_{\rm{T}}>$ 3~GeV/c, as shown in Figs. \ref{fig:openCharmRun3} and \ref{fig:openBeautyRun3}

\subsection{Low-mass dimuons}
\label{sub:lowMass}

\begin{figure}[h]
    \centering
    \begin{subfigure}[t]{0.5\linewidth}
        \centering
        \includegraphics[height=1.5in]{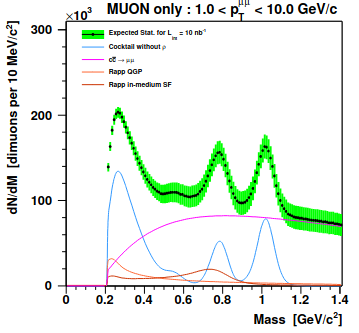}
        \caption{Run 2 performance}
        \label{fig:lowMassRun2}
    \end{subfigure}%
    ~ 
    \begin{subfigure}[t]{0.5\linewidth}
        \centering
        \includegraphics[height=1.5in]{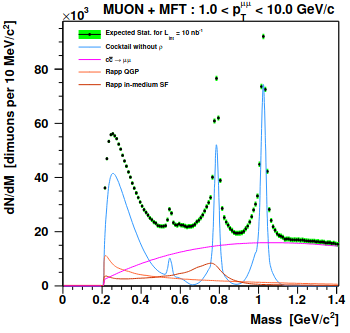}
        \caption{Expected Run 3 performance}
        \label{fig:lowMassRun3}
    \end{subfigure}
    \caption{Dimuon invariant-mass spectra in the low-mass region. Comparison between Run 2 and expected results in Run 3 (simulation) \cite{mftLOI}.}
\end{figure}

The dimuon invariant-mass is calculated as:

\begin{equation}
    \label{eq:invMass}
    \centering
    M_{\mu\mu} = \sqrt{2p_{1}p_{2}(1-cos \theta)} \approx \sqrt{p_{1}p_{2}\theta^{2}}
\end{equation}

where $p_{1}$ and $p_{2}$ are the momenta of the two muons and $\theta$ their opening angle. In the MS, the main contributor to the uncertainty on the invariant-mass, especially in the low-mass region, is related to the measurement of the dimuon opening angle, worsened by the multiple scattering caused by the hadron absorber. The installation of the MFT will enable the measurement of the opening angle upstream of the latter, improving both the S/B ratio as well as the invariant-mass resolution on the low-mass particles ($\sigma_{M}$). Figure \ref{fig:lowMassRun2} and \ref{fig:lowMassRun3} show the invariant-mass distribution in the low-mass region in Run 2 and Run 3 respectively. The S/B ratio and mass resolution improvements in Run 3 can clearly be observed and Table \ref{tab:massRes} reports the numerical values obtained with a Monte Carlo simulation. 

\begin{table}[h!]
\small
	\begin{center}
		\setlength{\tabcolsep}{5pt} 
        \begin{tabular}{ccc}\\\toprule  
        \textbf{Particle} & \textbf{Run 2 \textsigma $_{\text{M}}$ [MeV/$c^{2}$]} & \textbf{Run 3 \textsigma $_{\text{M}}$ [MeV/$c^{2}$]} \\\midrule
        $\eta$ & 44 & 8 \\  \midrule
        $\omega$ & 46 & 12 \\  \midrule
        $\phi$ & 51 & 15 \\  \bottomrule
        \end{tabular}
        \caption{Improvements of low-mass particles mass resolution with the upgraded MS in Run 3 \cite{mftLOI}.}\label{tab:massRes}
	\end{center}
\end{table}

\hspace{-2cm}
\section{Preliminary Run 3 results}
\label{sec:results}

\begin{figure}[h]
    \centering
    \begin{subfigure}[t]{0.5\linewidth}
        \centering
        \includegraphics[height=1.2in]{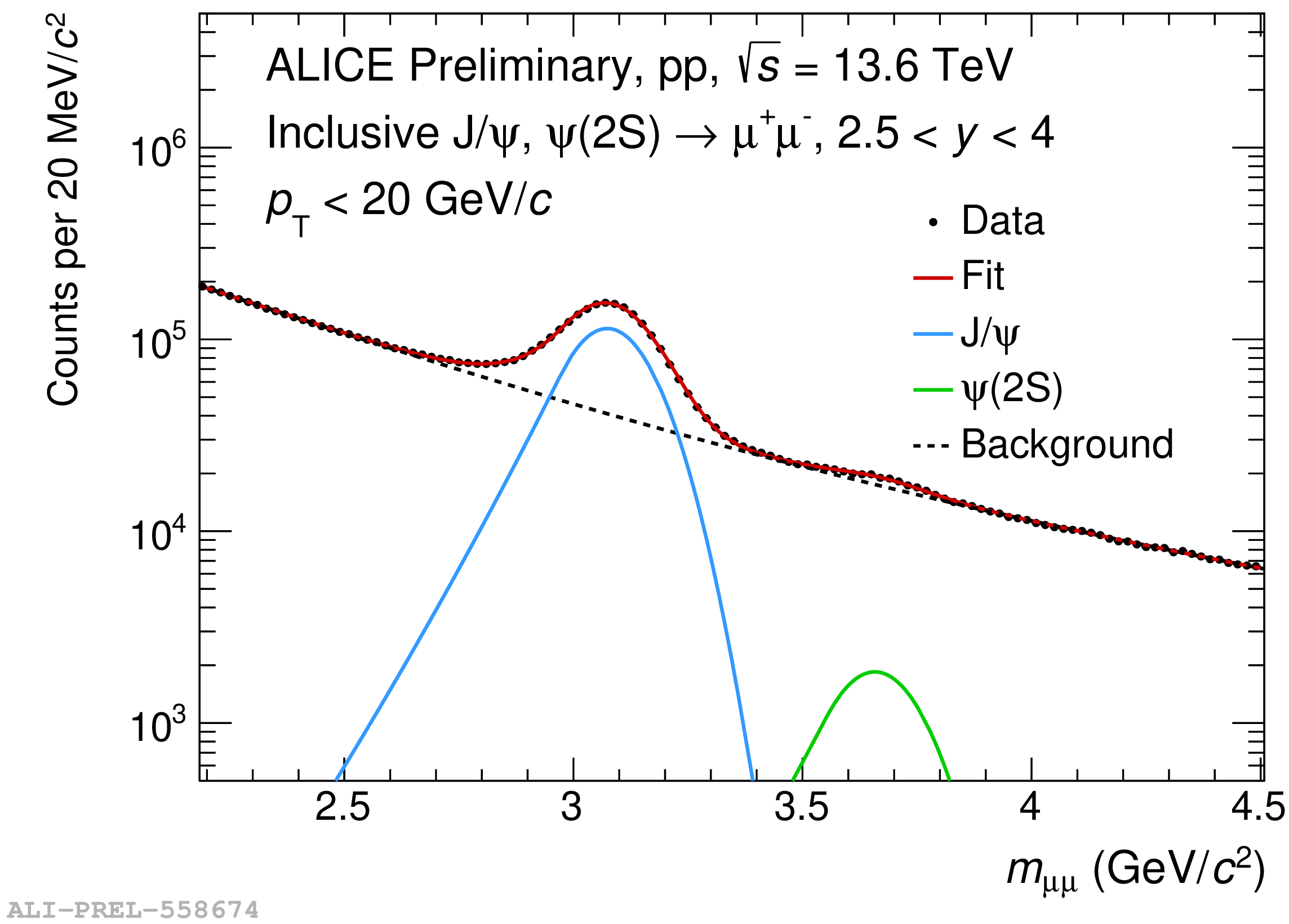}
        \caption{pp collisions}
        \label{fig:spectrumpp}
    \end{subfigure}%
    ~ 
    \begin{subfigure}[t]{0.5\linewidth}
        \centering
        \includegraphics[height=1.2in]{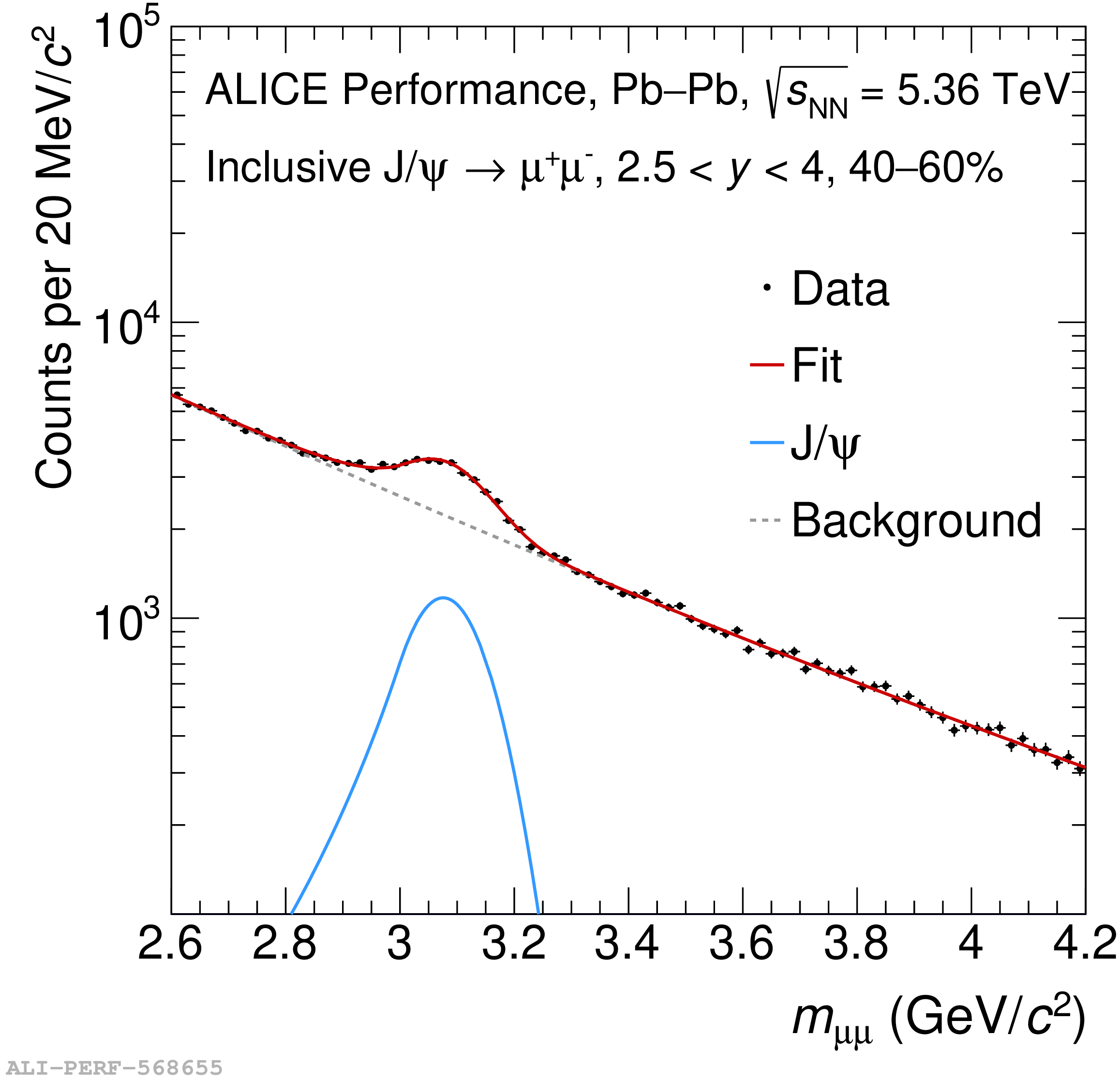}
        \caption{Pb\textendash{}Pb collisions}
        \label{fig:spectrumPbPb}
    \end{subfigure}
    \caption{Inclusive  J/$\psi$ and $\psi$(2S) production in pp collisions $at \sqrt{s}$ = 13.6~TeV (left) and Pb\textendash{}Pb collisions at $\sqrt{s_{\text{NN}}}$ = 5.36~TeV (right) in Run 3.}
\end{figure}

In this section, a selection of preliminary results, obtained with the MS using Run 3 data collected both in pp and Pb\textendash{}Pb collisions is shown. Figures \ref{fig:spectrumpp} and \ref{fig:spectrumPbPb} show the inclusive J/$\psi$ and $\psi$(2S) production obtained with he upgraded MS in pp and Pb\textendash{}Pb collisions at $\sqrt{s}$ = 13.6~TeV and $\sqrt{s_{\text{NN}}}$ = 5.36~TeV, respectively. 

\begin{figure}[h] 
\centering
\includegraphics[width=0.6\linewidth]{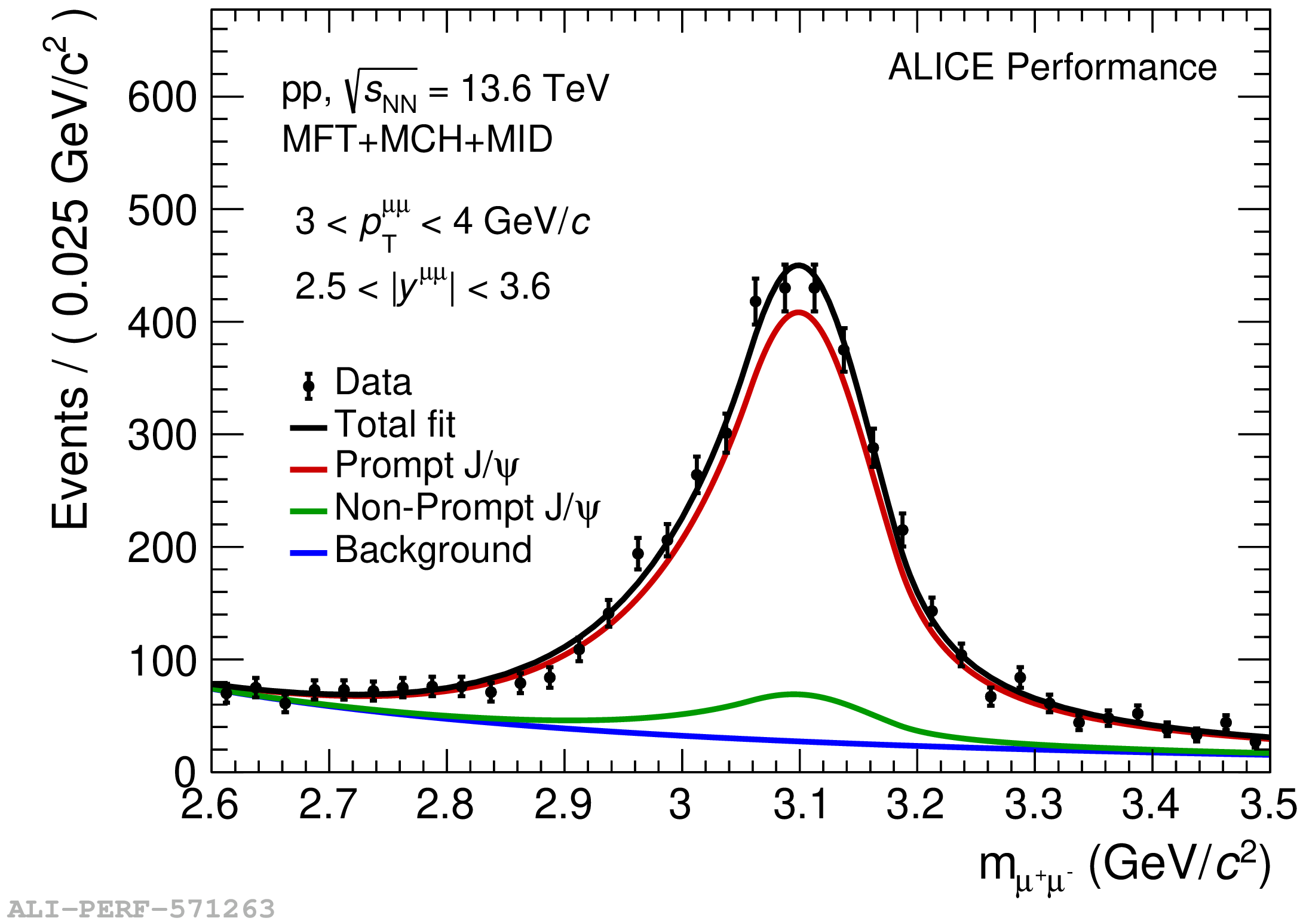}
\caption{Dimuon invariant-mass spectrum with the various fitted contributions extracted from a fit in pp collisions at $\sqrt{s}$ = 13.6~TeV.}
\label{fig:invMassDisplaced}
\end{figure}

Figure \ref{fig:invMassDisplaced} shows the dimuon invariant-mass spectrum obtained with the inclusion of the MFT. In particular, the deconvolution of the different contributions (prompt/non-prompt J/$\psi$ and background) are highlighted. These components have been obtained by following the procedure described in Section \ref{sub:nonPromptJpsi} and they will be used to calculate the ratio between the prompt and non-prompt J/$\psi$ at forward rapidity.

\section{Conclusions}
\label{sec:conclusion}

The ALICE MS has very well performed during the LHC Run 1 and 2, leading to many interesting results in the study of QGP in Pb\textendash{}Pb collisions at the LHC, some of which were summarized in this contribution.

The MS, together with the other ALICE subsystems, was upgraded in view of the LHC Run 3, in order to cope with the higher expected interaction rate and to work in continuous readout mode. Specifically, a new silicon pixel tracker, the MFT, was installed in front of the hadron absorber and it greatly enhances the capabilities of the MS by:

\begin{itemize}
    \item Increasing the S/B ratio for the J/$\psi$ and $\psi$(2S)

    \item Allowing to disentangle prompt and non-prompt J/$\psi$ at forward rapidity

    \item Allowing the study of open heavy-flavor separately in the single $\mu$ channel at forward rapidity

    \item Improving the S/B ratio and mass resolution for low-mass dimuon pairs
\end{itemize}

The MFT has been fully integrated in the ALICE data-taking and some of the preliminary results obtained from the analysis of Run 3 were discussed. These preliminary results are promising and further physics results are expected.

\bibliographystyle{unsrt}

\end{document}